\documentclass[twocolumn]{aastex631}  

\usepackage{graphicx}
\usepackage{txfonts}
\usepackage{natbib}
\begin{document}

\title{A Metal-Rich Atmosphere with a Super-Solar C/O Ratio for the Extreme Ultra-Hot Jupiter WASP-178b}

\author[0000-0001-8018-0264]{Suman Saha}
\affiliation{Instituto de Estudios Astrofísicos, Facultad de Ingeniería Ciencias, Universidad Diego Portales, Av. Ejército Libertador 441, Santiago, Chile}
\affiliation{Centro de Excelencia en Astrofísica y Tecnologías Afines (CATA), Camino El Observatorio 1515, Las Condes, Santiago, Chile}
	
\correspondingauthor{Suman Saha}
\email{suman.saha@mail.udp.cl}

\author[0000-0003-2733-8725]{James S. Jenkins}
\affiliation{Instituto de Estudios Astrofísicos, Facultad de Ingeniería Ciencias, Universidad Diego Portales, Av. Ejército Libertador 441, Santiago, Chile}
\affiliation{Centro de Excelencia en Astrofísica y Tecnologías Afines (CATA), Camino El Observatorio 1515, Las Condes, Santiago, Chile}

\begin{abstract}
The population of ultra-hot Jupiters (UHJs) provide unique opportunities to probe the extreme formation and evolutionary pathways in exoplanets. Owing to their very high temperatures and inflated atmospheres, UHJs are among the most favorable targets for both transmission and emission spectroscopy, enabling detailed characterization of their atmospheric properties. Here, we present a reanalysis of the JWST NIRSpec/G395H transmission spectra of the extreme ultra-hot Jupiter (EUHJ) WASP-178b, aimed at precisely characterizing its atmospheric composition. Our approach combines data reduction using two independent pipelines, lightcurve modeling with robust detrending techniques, and rigorous atmospheric retrievals. We report statistically significant detections of CO (7.24 $\sigma$) and CO$_2$ (7.22 $\sigma$), along with marginal evidence for C$_2$H$_2$ (1.34 $\sigma$), but no clear evidence for H$_2$O, suggesting depletion. From these retrieved abundances, we constrain the C/O ratio to a precise super-solar value of 0.954$\pm$0.033, consistent with an emerging trend in other UHJs. We also infer a very high atmospheric metallicity for a Jupiter-sized gas giant—11.44$_{-6.94}^{+12.54}$ $\times$solar—indicating unique atmospheric evolutions. These findings provide a critical benchmark for an extreme exoplanet atmosphere, offering a testbed for developing next-generation atmospheric evolution models and enabling comparative population-level studies across the UHJ population.
\end{abstract}

\section{Introduction}

Ultra-hot Jupiters (UHJs), defined by equilibrium temperatures exceeding $\sim$2000K, are emerging as one of the most important planetary populations for atmospheric characterization. Owing to their extreme physical and chemical conditions, they serve as natural laboratories for probing the boundaries of planet formation and evolutionary processes \citep[e.g.,][]{2011Natur.469...64M, 2018A&A...617A.110P, 2024ApJS..270...34C, 2024A&A...684A..27D, 2025A&A...700A..45S}. Transmission spectroscopy—and to a lesser extent emission spectroscopy— remain the most effective techniques for characterizing the atmospheres of transiting exoplanets \citep[e.g.,][]{2001ApJ...553.1006B, 2008ApJ...673L..87R, 2008ApJ...686.1341C, 2014ApJ...796..100T, 2024Natur.626..979P}. While many exoplanet populations remain beyond the reach of current instruments for precise atmospheric studies, UHJs stand out as particularly favorable targets for both transmission and emission spectroscopy. This is largely due to their extremely close-in orbits, high irradiation levels, and inflated atmospheres, which together enhance observational sensitivity.

While UHJs occupy the extreme ends of known exoplanet parameter space—particularly in equilibrium temperature and radius—no clear correlations have yet been identified among these properties, complicating efforts to infer their formation pathways. Moreover, UHJs are significantly rarer than the typical hot Jupiters, with only a small number of confirmed cases despite over two decades of intensive transit and radial velocity surveys. This rarity is especially striking given that the observational biases inherent to the transit method strongly favor the detection of UHJs over most other planet types. Together, these observations suggest that UHJs represent a fundamentally distinct class of exoplanets, likely formed under unique conditions that cannot be easily generalized to broader exoplanet populations—or even consistently among UHJs themselves \citep[e.g.,][]{2019A&A...626A.133H, 2021ApJ...907L..22C, 2023ApJ...951..123H}.

\begin{figure}
\centering
\includegraphics[width=\columnwidth]{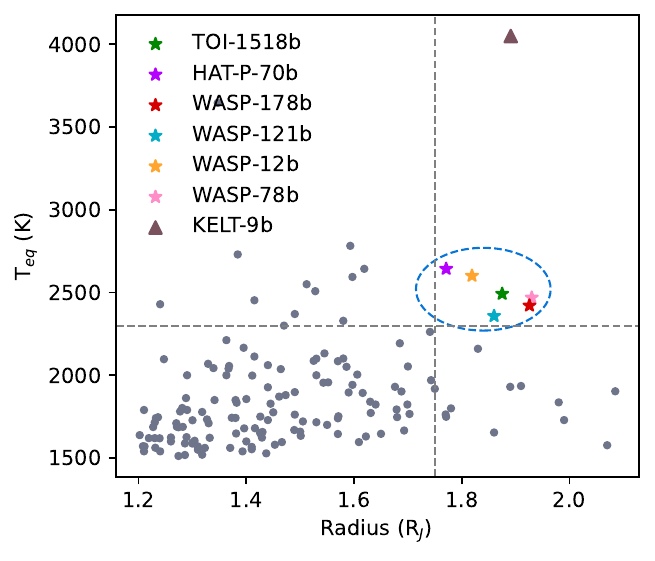}
\caption{Planetary radius versus equilibrium temperature for known hot and ultra-hot Jupiters. Extreme ultra-hot Jupiters (EUHJs) are highlighted as colored stars within the blue dashed ellipse, highlighting a distinct population. KELT-9b is marked as a brown triangle. Dashed black lines indicate approximate boundaries separating EUHJs from the broader population. \label{fig:fig1}}
\end{figure}

By comparing key parameters of known UHJs—such as equilibrium temperature and radius (see Figure \ref{fig:fig1})—we identify a distinct subset that simultaneously exhibits exceptionally high temperatures and large radii. These planets, highlighted by the colored stars within the blue dashed ellipse, stand apart from the broader UHJ population, which typically displays only one of these extreme properties or neither. We designate this subset as extreme ultra-hot Jupiters (EUHJs). This classification is particularly relevant to our study, as our target planet, WASP-178b \citep{2019MNRAS.490.1479H}, falls within this group, which currently includes only six known members. Notably, KELT-9b \citep{2017Natur.546..514G}, despite its extreme temperature, is excluded from this category due to its distinct characteristics and outlier status within this parameter space.

WASP-178b has emerged as one of the most compelling targets for atmospheric characterization since its discovery. As an ultra-hot Jupiter with a highly inflated atmosphere (Mp = 1.66 M$_J$ and Rp = 1.87 R$_J$, \citealt{2023ApJS..268....2S}) orbiting a bright A-type star (V $\approx$ 10), it is particularly well suited for both transmission and emission spectroscopic studies. Transmission spectroscopy with HST previously suggested the presence of SiO in its atmosphere \citep{2022Natur.604...49L}, representing the first proposed detection of this molecule in an exoplanet. In addition, high-resolution ground-based transmission spectroscopy with VLT/ESPRESSO has revealed several atomic absorption features, such as Na I, H$\alpha$, and H$\beta$, etc \citep{2024A&A...689A..54D}. Ground-based emission spectroscopy using VLT/CRIRES+ \citep{2024A&A...688A.206C} has also detected CO and H$_2$O, and suggested a super-solar metallicity with a possibly solar to super-solar C/O ratio, albeit with significant uncertainty. 

More recently, \cite{2025AJ....169..274L} (hereafter L25) analyzed the James Webb Space Telescope (JWST) \citep{2006SSRv..123..485G} NIRSpec/G395H transmission spectra of WASP-178b. They reported muted H$_2$O and CO features, along with non-detection of CO$_2$ and SiO, but without providing statistical evidence for either detection or non-detection. They also inferred a loosely constrained sub-solar C/O ratio, in contrast to the recent reports of super-solar C/O ratios in several UHJ atmospheres \citep[e.g.,][]{saha2025a, saha2025b}. These findings highlight the need for an independent analysis of the JWST dataset to verify the earlier results and potentially reveal deeper insights into the atmospheric conditions of this uniquely valuable EUHJ.

The carbon-to-oxygen (C/O) ratio is widely regarded as a key tracer of the formation and migration history of giant exoplanets \citep{2011ApJ...743L..16O, 2018A&A...613A..14E, 2019A&A...632A..63C}. Super-solar C/O ratios inferred for several UHJs \citep{2011Natur.469...64M, 2024AJ....168...14W, 2024AJ....168..293S} have been interpreted as evidence for formation beyond the snow line, followed by inward migration. However, the relatively large uncertainties in most current measurements limit their ability to robustly constrain such formation pathways. The C/O ratio has also been proposed as a diagnostic for distinguishing between formation via core accretion and gravitational instability \citep{2014ApJ...794L..12M, 2024ApJ...969L..21B}. In addition, oxygen depletion through sequestration into refractory condensates has been suggested as a mechanism that could artificially elevate the observed C/O ratios in gas giant atmospheres \citep{2023MNRAS.520.4683F}.

Robust estimations of the C/O ratio in planetary atmospheres require precise measurements of the abundances of both carbon- and oxygen-bearing molecules. Many of these species exhibit strong absorption features in the near- to mid-infrared wavelengths where earlier ground-based instruments, limited in both sensitivity and wavelength coverage, have struggled to achieve the necessary precision. JWST, with its broad wavelength coverage and high-precision spectroscopic capabilities, represents a significant advancement in this regard. It enables the simultaneous detection of multiple key molecular species, allowing for unprecedented accuracy in the retrieval of atmospheric abundances and C/O ratios. These measurements also provide insights into planetary metallicities, which serve as important tracers of atmospheric mass loss through photoevaporation \citep[e.g.,][]{2018ApJ...865...75N, 2020arXiv200508676M} and of disk accretion histories shaped by planetary migration \citep[e.g.,][]{1998AREPS..26...53B, 2021MNRAS.503.5254L, 2023A&A...675A.176U}.

In this work, we present a reanalysis of the JWST NIRSpec/G395H transmission spectroscopy of WASP-178b, aiming to accurately constrain the abundances of key atmospheric constituents, as well as assess the potential presence of clouds. Building on these tightly constrained molecular abundances, we further estimate the planet's C/O ratio and metallicity with high precision. These measurements will facilitate comparative studies of fundamental parameters across planetary populations, potentially revealing the dominant processes driving atmospheric evolution in UHJs and beyond. In Section \ref{sec:sec2}, we detail our methodology, covering data reduction, lightcurve analysis, and atmospheric retrievals. In Section \ref{sec:sec3}, we present the key findings from this work and discuss their broader implication in detail.

\section{Methodology}\label{sec:sec2}

\subsection{Observational data}\label{sec:sec2.1}

WASP-178b was observed by JWST during Cycle 1 as part of the GO program \#2055 (PI: J Lothringer). The observation took place on March 7, 2023, using the Near-Infrared Spectrograph \citep[NIRSpec,][]{2022A&A...661A..80J} in the Bright Object Time Series \citep[BOTS, e.g.,][]{2023PASP..135a8002E} mode, employing the G395H/F290LP disperser/filter configuration. This setup allows mid- to high-resolution spectroscopy over a broad wavelength range in near-infrared, approximately 2.8-5.2 $\mu$m.

We accessed the publicly available archival data via the Barbara A. Mikulski Archive for Space Telescopes (MAST)\footnote{https://mast.stsci.edu/} \dataset[(doi: 10.17909/wz25-4465)]{\doi{10.17909/wz25-4465}}. A check of the timing information confirmed that the observation spanned approximately seven hours in total, encompassing the full transit event of WASP-178b with sufficient out-of-transit baselines both before and after the event. This complete coverage is essential for accurate lightcurve detrending, which is particularly important in high-precision studies such as transmission spectroscopy, as incomplete coverage could lead to inaccuracies in lightcurve modeling and biases in the resulting spectra.

\subsection{Data reduction and analyses}\label{sec:sec2.2}

The JWST data available from MAST includes both raw files and products from various stages of processing by the official JWST Science Calibration Pipeline \citep{2022zndo...7071140B}. The raw files (denoted as “\texttt{*uncal.fits}") are four-dimensional, comprising the detector's column and row axes, the number of groups per integration, and the number of integrations per exposure. The dataset includes observations from both NIRSpec detectors, NRS1 and NRS2, with data from each detector further divided into multiple sectors (three in this case) to manage the large data volume.

We reduced the raw data files using two independent custom pipelines available from the community, namely \texttt{Eureka!} \citep{Bell2022} and \texttt{exoTEDRF} \citep[formerly \texttt{supreme-SPOON},][]{2024JOSS....9.6898R}. Both pipelines are well-adopted and have been rigorously tested for accuracy (especially \texttt{Eureka!}) through several previously published works (e.g., \citealt{2023Natur.614..664A, 2023NatAs...7.1317L, 2024Natur.626..979P, 2023ApJ...959L...9M} for \texttt{Eureka!}; and \citealt{2024ApJ...962L..20R, 2023Natur.614..670F} for \texttt{exoTEDRF}). In addition to the science frames, these reduction pipelines also require calibration files, which were accessed via the JWST Calibration Reference Data System (CRDS) \footnote{https://jwst-crds.stsci.edu/}. Using two independent pipelines enables us to cross-check the robustness of our reduction and analyses by comparing the results—an important step in high-precision studies such as this.

\subsubsection{\texttt{Eureka!} reduction}

\texttt{Eureka!} provides a total of six stages for the data reduction and analysis of JWST time-series data across various instruments and observing modes, each consisting of a set of built-in functions and sub-processes \citep{Bell2022}. For this work, we adopted the first three stages, which include calibration, correction procedures, and optimal extraction of the spectroscopic data. The NRS1 and NRS2 datasets were reduced separately. We largely followed the default configurations provided by the custom “\texttt{*.ecf}" files for NIRSpec/G395 spectral reduction, applying only minor modifications to optimize performance.

The jump rejection threshold was set to 8$\sigma$, as higher thresholds have been recommended for time-series analysis. For background subtraction, we used the “median" method rather than “mean" to avoid the vertical drift otherwise observed in the extracted time-series data at certain pixels. The half-width of the extraction aperture was set to 6 pixels, and we found that the SNR remained stable when the aperture size was varied slightly around this value. Among other correction stages, \texttt{Eureka!} stage 2 also includes 1/f noise correction, a crucial process highlighted in several previous studies \citep[e.g.][]{2023PASP..135a8002E, 2016SPIE.9910E..16P, 2022SPIE12180E..2PB}.

\subsubsection{\texttt{exoTEDRF} reduction}

\texttt{exoTEDRF} also provides a streamlined framework for reducing raw JWST data across multiple instrumental modes, organized into four processing stages. For our analysis, we adopted the first three stages, which include calibration, correction, and spectral extraction. As with \texttt{Eureka!}, the NRS1 and NRS2 datasets were reduced independently. We largely followed the default configuration setting tailored for NIRSpec/G395H, applying only minimal adjustments. Specifically, the extraction box width was set to 8 pixels; slight variations in this parameter did not produce any significant changes in the extracted spectra. Similar to \texttt{Eureka!}, \texttt{exoTEDRF} implements the 1/f noise correction step, which was also included in our analysis.

\begin{figure*}
\centering
\includegraphics[width=2\columnwidth]{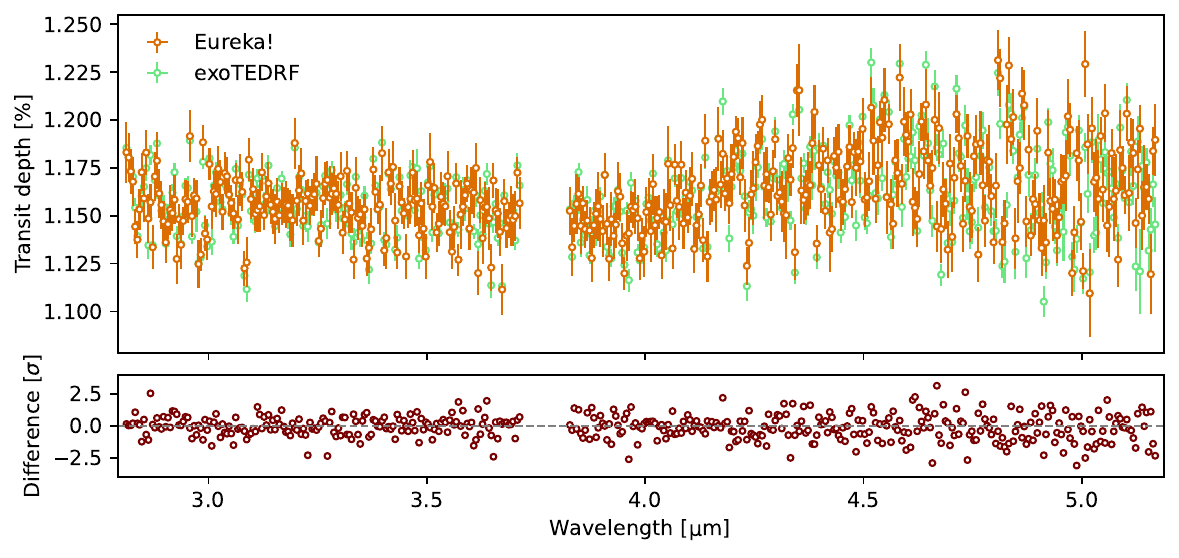}
\caption{Observed transmission spectra of WASP-178b from NIRSpec/G395H, extracted with two independent reduction pipelines—\texttt{Eureka!} (orange) and \texttt{exoTEDRF} (green). The lower panel shows their difference in units of \texttt{Eureka!} incertainties. The mean absolute difference is $\sim$1$\sigma$, indicating excellent statistical agreement. \label{fig:fig2}}
\end{figure*}

\subsubsection{Lightcurve modeling}\label{sec:sec2.3}

To obtain the transmission spectra from the time-series spectroscopic data extracted by the reduction pipelines (i.e., the 2D lightcurves), we binned the data along the wavelength axis. After testing several binning schemes, we adopted a fine 0.005 $\mu$m bin width, corresponding to a spectral resolution of approximately 1200-2000. This fine binning was feasible due to the high spectral resolution of NIRSpec/G395H and the excellent signal-to-noise ratio (SNR) of the observations. The binning produced 450 spectroscopic lightcurves—181 from NRS1 coverage, and 269 from NRS2 coverage (see Figure \ref{fig:figA1}). We discarded the first 45 exposures due to significant systematic noise, possibly caused by instrument settling. This exclusion does not affect our lightcurve modeling, as a substantial out-of-transit baseline (more than 1.5 hours) was observed prior to ingress.

The transit lightcurves were fitted using our in-house developed pipeline, \texttt{ExoELF} (ExoplanEts Lightcurves Fitter) \citep{2025MNRAS.539..928S, saha2025a, saha2025b}, which integrates several widely adopted packages. These include \texttt{batman} \citep{2015PASP..127.1161K} for transit and eclipse modeling, \texttt{emcee} \citep{2013PASP..125..306F} and \texttt{DYNESTY} \citep{2020MNRAS.493.3132S} for MCMC and nested sampling, respectively, and \texttt{celerite} \citep{celerite1, celerite2} and \texttt{george} \citep{2015ITPAM..38..252A} for Gaussian process (GP) regression. The pipeline is optimized to incorporate various detrending schemes and to efficiently handle large ensembles of lightcurves, such as those obtained from the time-series spectroscopic observations with JWST.

At first glance, the binned lightcurves from the \texttt{exoTEDRF} reduction appeared to have smaller uncertainties; for example, the mean uncertainty in the first binned lightcurve was less than half that of the \texttt{Eureka!} reduction. However, upon closer inspection, we found that the uncertainties in the \texttt{exoTEDRF} lightcurves were constant across pixels, despite exhibiting large fluctuations between them (see Figure \ref{fig:figAerr}). This behavior likely stems from computational approximations during the data reduction process, resulting in an underestimation of the uncertainty levels. When comparing the white lightcurves, we observed that these fluctuations dominate the \texttt{exoTEDRF} uncertainty estimates (see Figure \ref{fig:figAerr}), reducing the apparent discrepancy between the two pipelines.

Consequently, we consider the \texttt{Eureka!} reduction to produce more internally consistent uncertainty estimates, and have adopted it as the default for our analyses. Nonetheless, we performed all analysis steps using both datasets and compared the results to assess the robustness of our findings. This approach of selecting the best-performing reduction as the default has been widely adopted in similar studies \citep[e.g.][]{2023NatAs...7.1317L, 2023Natur.614..653A, 2023Natur.614..670F}. While some works have shown that different pipelines can yield divergent results \citep[e.g.][]{2023ApJ...943L..10C}, in our case, we find consistent results across reductions, as discussed later.

We have adopted a two-stage procedure for extracting the transmission spectra from the spectroscopic lightcurves. In the first stage, we modeled the white lightcurve from NRS1 and NRS2 jointly, allowing all the model parameters to vary freely. In the second stage, we modeled each spectroscopic lightcurve independently, fixing the wavelength-independent parameters to the best-fit values obtained from the white lightcurve analysis.

The free parameters in the white lightcurve modeling included the mid-transit time, scaled orbital semi-major axis, orbital inclination, planet-to-star radius ratio, quadratic limb darkening coefficients, and detrending terms (see Table \ref{tab:tr_fit}, Figure \ref{fig:fig_lc_w}, \ref{fig:figA1b}). The orbital period and orbital eccentricity were held fixed. In the spectroscopic lightcurve modeling, we further fixed the mid-transit time, scaled orbital semi-major axis, and inclination, while fitting for the planet radius and wavelength-dependent detrending parameters.

\begin{figure*}
\centering
\includegraphics[width=2\columnwidth]{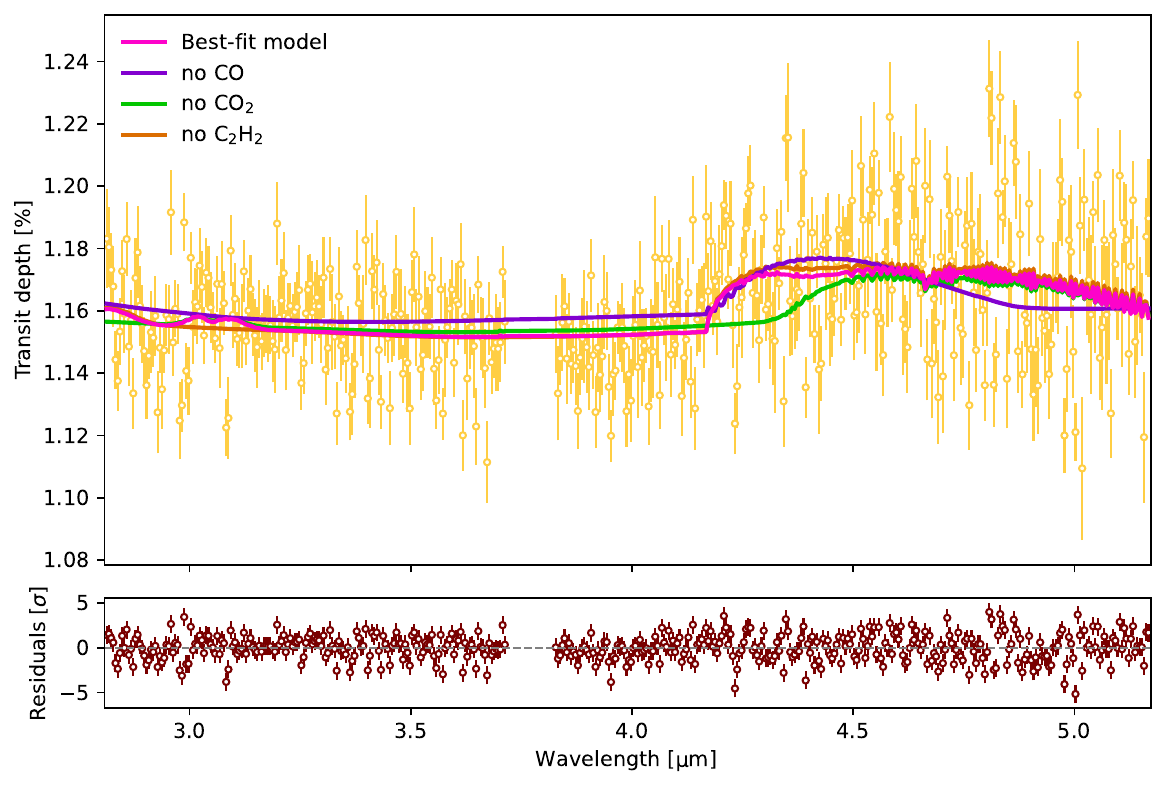}
\caption{Observed transmission spectrum of WASP-178b (yellow, \texttt{Eureka!}) with the best-fit model from free chemistry retrieval. Residuals are shown in units of observational uncertainties. Additional retrievals excluding CO, CO$_2$, and C$_2$H$_2$ are overplotted, which were used to assess the statistical significance of their detection.\label{fig:fig3}}
\end{figure*}

We extensively tested several detrending techniques, including linear and quadratic polynomials and GP regression, to model the systematics in the lightcurves. This step is crucial for high-sensitivity analyses, as even subtle trends in the transmission spectra can introduce significant biases. Although the lightcurves appeared mostly flat with only minor long-term trends, polynomial detrending tended to overfit the data, resulting in overestimated transit depths (see Figure \ref{fig:figdetren}). In contrast, GP regression provided a more robust treatment of the systematics, yielding transit depth estimates with significantly higher precision while remaining statistically consistent with other methods. Our GP implementation used the \texttt{celerite} package with a Matérn-3/2 kernel. The joint lightcurve fitting was performed using nested sampling via \texttt{DYNESTY}.

\begin{figure*}
\centering
\begin{tabular}{c c}
\multicolumn{2}{c}{\includegraphics[width=\columnwidth]{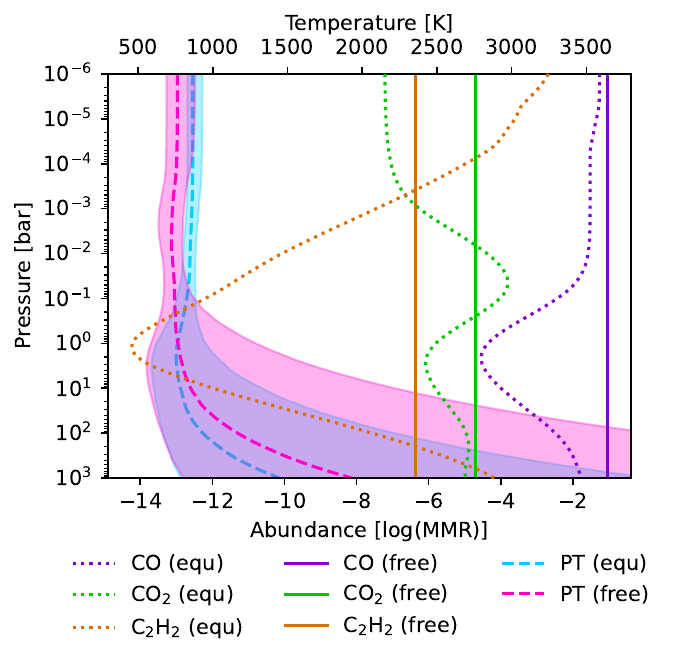}} \\[2mm]
\includegraphics[width=\columnwidth]{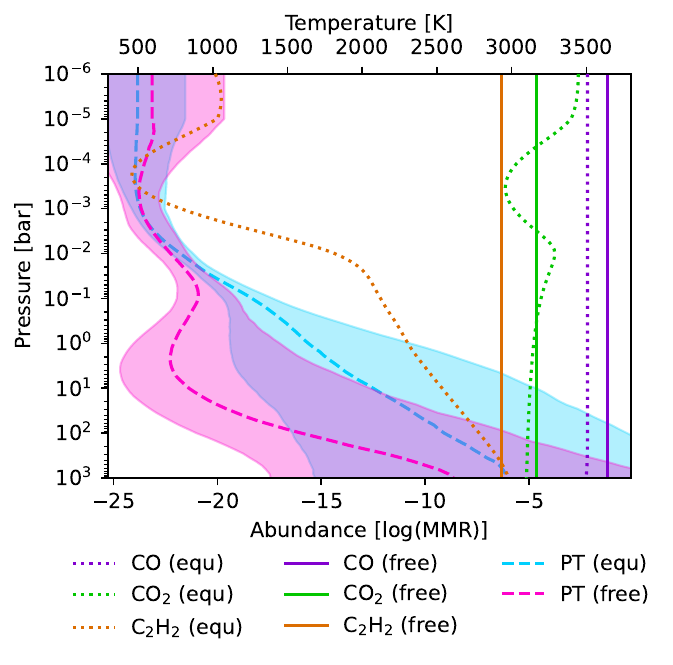} &
\includegraphics[width=\columnwidth]{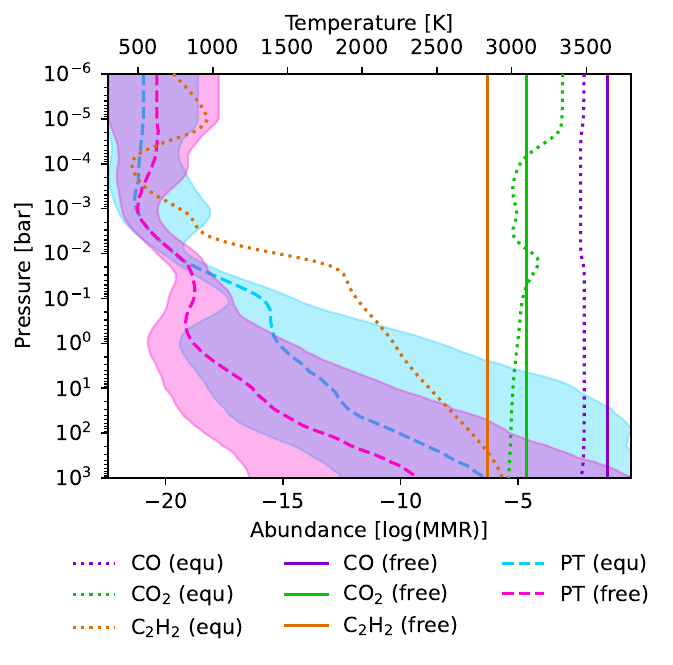} \\
\end{tabular}
\caption{Pressure-temperature (PT) profiles from the equilibrium and free-chemistry retrievals using Guillot (top), dtdp6 (bottom left), and dtdp10 (bottom right) parametrization, together with the retrieved median abundances of CO, CO$_2$, and C$_2$H$_2$. \label{fig:fig4}}
\end{figure*}

The resulting transmission spectra from both independent data reduction procedures are shown in Figure \ref{fig:fig2}. The mean absolute difference between the two spectra is approximately 1$\sigma$, indicating excellent statistical agreement. While the spectra are visually similar, the data points from the \texttt{exoTEDRF} reduction exhibit smaller formal uncertainties, reflecting the smaller per-point uncertainties in the binned lightcurves discussed earlier. However, this does not correspond to a lower overall scatter of the data; in fact, the \texttt{exoTEDRF} spectrum has a variance approximately $10\%$ higher than the \texttt{Eureka!} spectrum.

\begin{table*}
    \centering
    \caption{Molecular abundances (in log-MMRs) derived from the retrieval analyses.}
    \label{tab:eureka_free}
    \[
    \begin{array}{l l l c c c c}
        \hline
        \hline
        \text{Data} & \text{PT profile} & \text{Model} & [\mathrm{CO}] & [\mathrm{CO}_2] & [\mathrm{C}_2\mathrm{H}_2] & \ln Z \\
        \hline
        \text{Eureka!} & \text{Guillot} & \text{free} &
        -1.04_{-0.38}^{+0.28} & -4.69_{-0.35}^{+0.38} & -6.37_{-1.95}^{+0.70} & 3137.25 \\
        & & \text{free (no CO)} &
        \text{---} & -7.12_{-0.14}^{+0.18} & -7.05_{-3.15}^{+1.69} & 3111.07 \\
        & & \text{free (no CO$_2$)} &
        -0.52_{-0.42}^{+0.17} & \text{---} & -6.43_{-2.27}^{+0.84} & 3111.17 \\
        & & \text{free (no C$_2$H$_2$)} &
        -1.12_{-0.42}^{+0.31} & -4.75_{-0.40}^{+0.37} & \text{---} & 3136.36 \\
        & \text{dtdp6} & \text{free} &
        -1.23_{-0.38}^{+0.33} & -4.65_{-0.35}^{+0.38} & -6.33_{-1.67}^{+0.70} & 3137.26 \\
        & \text{dtdp10} & \text{free} &
        -1.20_{-0.39}^{+0.31} & -4.66_{-0.34}^{+0.35} & -6.29_{-1.37}^{+0.64} & 3138.31 \\
        & \text{Guillot} & \text{equilibrium} &
        -1.76_{-0.23}^{+0.22} & -5.03_{-0.42}^{+0.45} & -4.44_{-1.41}^{+0.86} & 3135.78 \\
        & \text{dtdp6} & \text{equilibrium} &
        -2.26_{-0.17}^{+0.25} & -3.81_{-0.54}^{+0.48} & \text{---} & 3137.17 \\
        & \text{dtdp10} & \text{equilibrium} &
        -2.31_{-0.31}^{+0.29} & -4.30_{-0.72}^{+0.68} & -7.07_{-0.56}^{+0.33} & 3137.54 \\
        \text{exoTEDRF} & \text{Guillot} & \text{free} &
        -0.93_{-0.23}^{+0.18} & -4.39_{-0.23}^{+0.25} & -6.74_{-0.77}^{+0.40} & 1768.67 \\
        & \text{dtdp6} & \text{free} &
        -1.18_{-0.24}^{+0.22} & -4.42_{-0.22}^{+0.24} & -6.79_{-0.87}^{+0.41} & 1773.05 \\
        & \text{dtdp10} & \text{free} &
        -1.17_{-0.25}^{+0.22} & -4.43_{-0.22}^{+0.23} & -6.79_{-0.73}^{+0.4} & 1773.73 \\
        & \text{Guillot} & \text{equilibrium} &
        -1.46_{-0.19}^{+0.16} & -4.40_{-0.69}^{+0.40} & -4.50_{-0.28}^{+1.54} & 1773.85 \\
        & \text{dtdp6} & \text{equilibrium} &
        -3.06_{-0.20}^{+0.33} & -3.89_{-0.79}^{+0.60} & \text{---} & 1771.36 \\
        & \text{dtdp10} & \text{equilibrium} &
        -2.54_{-0.3}^{+0.12} & -4.35_{-0.62}^{+0.25} & \text{---} & 1773.37 \\
        \hline
        \hline
    \end{array}
    \]
\end{table*}

\subsection{Atmospheric retrieval analyses}\label{sec:sec2.4}

The combination of high spectral resolution and excellent SNR provided by the JWST observations enables detailed atmospheric retrieval analyses of the transmission spectra using state-of-the-art modeling frameworks. For this purpose, we adopted \texttt{petitRADTRANS} \citep[\texttt{pRT},][]{2019A&A...627A..67M}, a widely used and well-validated radiative transfer code within the exoplanet community \citep[e.g.,][]{2024arXiv241008116K, 2024ApJ...973L..41I, 2024arXiv241203675L, 2024MNRAS.527.7079P, 2024A&A...690A..63B}. Although pRT employs a one-dimensional radiative transfer formalism, it remains sufficiently robust for the reliable interpretation of exoplanet spectra observed with current instrumentation, and it provides nearly all required functionalities for such complex analyses.

We built our retrieval framework based on the \texttt{pRT} retrieval routines \citep{2024JOSS....9.5875N}, which integrate \texttt{PyMultinest} \citep{2009MNRAS.398.1601F, 2014A&A...564A.125B} for nested sampling of the posterior space. Given the high spectral resolution of our extracted transmission spectra, we used the line-by-line opacity mode. Specifically, we set the model resolution to 6000, which is well-suited to the spectral resolution of our data. 

The pressure-temperature (PT) profiles play a critical role in atmospheric retrievals. We adopted the physically motivated parametrization of \citet{2010A&A...520A..27G} (hereafter Guillot), which is well-suited for low-resolution retrievals, providing sufficient flexibility without unnecessary computational complexity. The parametrization is governed by four variables—the equilibrium (upper-atmosphere) temperature (T$\mathrm{_{eq}}$), the internal temperature (T$\mathrm{_{int}}$), the visible-to-infrared opacity ratio ($\gamma$), and the mean infrared opacity ($\kappa_{\mathrm{IR}}$). To provide a comparative abundance estimation and validate the results obtained with the Guillot PT profile, we also employed the more flexible parameterization by \citet{2023AJ....166..198Z}, using 6 spline points (dtdp6) and 10 spline points (dtdp10). In these profiles, the temperature gradients at a fixed number of spline points (6 or 10) are treated as free parameters (s$_{PT(1-10}$) along with the internal temperature (T$_{bottom}$).

We employed both free and equilibrium chemistry models to define the abundances of chemical species in our retrieval analyses. In the free chemistry approach, the abundances (expressed as log-MMRs) of individual species were treated as vertically uniform free parameters, typically ranging between -12 and -0.2. To improve computational efficiency, the upper bound was lowered for some species that did not show evidence of requiring a higher range in preliminary fits. For the equilibrium chemistry models, we used precomputed abundance tables provided by \texttt{pRT}, assuming chemical equilibrium at each pressure level. We included the opacities of all key molecules expected to contribute significantly to the transmission spectra across the observed wavelengths, including H$_2$O \citep{2018MNRAS.480.2597P}, CO \citep{2010JQSRT.111.2139R}, CO$_2$ \citep{2010JQSRT.111.2139R}, SiO \citep{2013MNRAS.434.1469B}, CH$_4$ \citep{2020ApJS..247...55H}, HCN \citep{2006MNRAS.367..400H}, C$_2$H$_2$ \citep{2013JQSRT.130....4R}, PH$_3$ \citep{2015MNRAS.446.2337S}, and H$_2$S \citep{2013JQSRT.130....4R}. Rayleigh scattering by H$_2$ \citep{1962ApJ...136..690D} and He \citep{1965PPS....85..227C} was included, along with collision-induced absorption (CIA) from H$_2$-H$_2$ \citep{2001JQSRT..68..235B, 2002A&A...390..779B} and H$_2$-He \citep{1988ApJ...326..509B, 1989ApJ...336..495B}. Additionally, bound-free and free-free absorption H$^-$ \citep{2008oasp.book.....G} were incorporated into our equilibrium chemistry model, as these processes can be significant at the high temperatures typical of UHJ atmospheres. We also included several condensate species expected to form clouds at these temperatures, such as Fe(s) \citep{2018MNRAS.475...94K, 1994ApJ...421..615P, 1997A&A...327..743H, 2001ApJ...556..872A}, MgSiO$_3$(s) \citep{1994A&A...292..641J, 2001ApJ...556..872A}, Al$_2$O$_3$(s) \citep{1995Icar..114..203K, Stull1947}, SiO$_2$(s) \citep{2018MNRAS.475...94K, 1997A&A...327..743H, 1995Icar..114..203K, Stull1947}, SiO(s) \citep{2013A&A...555A.119G}, TiO$_2$(s) \citep{2018MNRAS.475...94K, 2011A&A...526A..68Z, 2003ApJS..149..437P, 2016arXiv160704866S, 2019RJPCA..93.1024S}, and CaTiO$_3$(s) \citep{2018MNRAS.475...94K, 2003ApJS..149..437P, 1998JPCM...10.3669U, 2019RJPCA..93.1024S}. The abundances of these cloud species were treated as free parameters in both free and equilibrium chemistry models. To characterize the cloud properties, we included the sedimentation efficiency  (f$_{\mathrm{sed}}$), log-normal width of particle size distribution ($\sigma$), the vertical eddy diffusion coefficient (k$_{\mathrm{zz}}$), and the cloud fraction (f$_{\mathrm{c}}$) as additional free parameters. To account for the small offset between NRS1 and NRS2 data, we included an NRS2 scale factor ($f_{\mathrm{s, NRS2}}$) as a free parameter in the retrievals (also see Table \ref{tab:retrieval_priors}).

\section{Results and Discussion}\label{sec:sec3}

The high S/N spectroscopic coverage of JWST between 2.8 and 5.2 µm, combined with the highly inflated radius of WASP-178b, provides an ideal case for tightly constraining transmission spectroscopic retrievals. Free chemistry retrievals of the \texttt{Eureka!} spectra using the Guillot PT profile yielded well-constrained abundances of CO and CO$_2$, along with a marginal constraint on C$_2$H$_2$ (see Table \ref{tab:eureka_free}, Figure \ref{fig:figA_cor_free}). Contrary to expectations for this wavelength range, the retrievals did not constrain the abundances for H$_2$O and SiO (Figure \ref{fig:figA_cor_free_all}). In particular, the 2.8-3.5 µm region of the transmission spectra is notably flat, explaining the unconstrained H$_2$O abundance.

We also performed retrievals excluding each species to assess the statistical significance of their detection, finding significant detection of CO (7.24 $\sigma$) and CO$_2$ (7.22 $\sigma$), a marginal detection of C$_2$H$_2$ (1.34 $\sigma$), and no evidence for H$_2$O or SiO. Additionally, we found no statistical evidence for the presence of clouds. While clouds have been detected in other ultra-hot Jupiters, such as WASP-19b \citep{saha2025a} and WASP-121b \citep{saha2025b}, those studies included emission spectroscopy at shorter optical-to-NIR wavelengths (down to 0.6 µm), which are more sensitive to cloud signature. Therefore, the presence of clouds in WASP-178b cannot be ruled out based on this analysis.

Using the more flexible dtdp6 and dtdp10 PT profiles in free chemistry retrievals of the \texttt{Eureka!} spectra also yielded abundances of CO, CO$_2$ and C$_2$H$_2$ consistent within 1 $\sigma$ of the Guillot PT results (see Table \ref{tab:eureka_free}, Figure \ref{fig:fig4}). Similarly, free chemistry retrievals of the \texttt{exoTEDRF} spectra with all three PT profiles produced consistent results (see Table \ref{tab:eureka_free}), but with slightly smaller uncertainties owing to the higher precision of the spectrum, thereby providing further validation of the estimated abundances.

Equilibrium chemistry retrievals considering the statistically detected molecular species—CO, CO$_2$, and C$_2$H$_2$—produced fits with similar statistical evidence, though the estimated abundances were slightly different (see Table \ref{tab:eureka_free}). For example, the retrieved CO abundance was generally lower, which we attribute to the model constraining the CO/CO$_2$ abundance ratio. This arises because a small change in CO$_2$ abundance produces a much large spectral feature than a comparable change in CO. For a similar reason, the retrieved C$_2$H$_2$ abundance reached the lower boundary of the equilibrium chemistry tables in a few cases, which are left unfilled in Table \ref{tab:eureka_free}.

We also tested including H$_2$O into these equilibrium chemistry retrievals, but this resulted in non-physical high temperatures, skewed abundances for other detected species, and a much lower statistical likelihood of the fit. The cause is equilibrium chemistry tables did not allow for sufficiently low H$_2$O abundance to reproduce the muted spectral feature, forcing the model to compensate with unrealistically high temperature fits.

\begin{figure*}
\centering
\includegraphics[width=1.56\columnwidth]{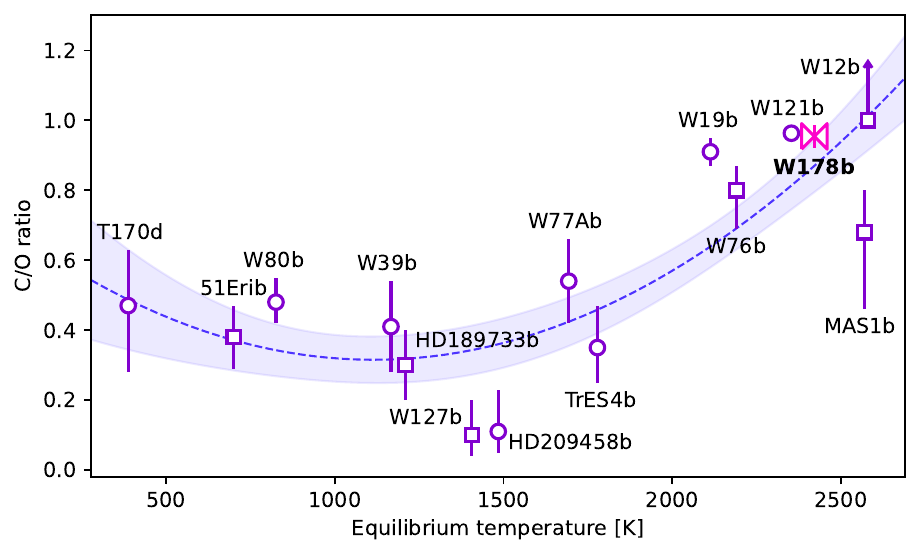}
\caption{Currently known and well-constrained C/O ratios for several exoplanets plotted against their equilibrium temperatures, including the estimate for WASP-178b from this work (magenta bowtie). Literature measurements based on JWST observations are shown as circles, while others are shown as squares. The distribution of C/O ratios is modeled as a quadratic function of equilibrium temperature, with the dashed blue line indicating the mean trend and the shaded area the 1$\sigma$ confidence interval. \label{fig:figco}}
\end{figure*}

In contrast to L25, our strong statistical detection of both CO and CO$_2$, along with other differences, warrant further discussion. Focusing on the abundances from the free chemistry retrievals (since we already established the limitations of the equilibrium chemistry retrievals), we none that L25 reported precise constraints on H$_2$O and SiO, despite claiming no detection of SiO and muted features for H$_2$O. In general, non-detections in retrievals should result in unconstrained lower bounds for the species in question (e.g., see Figure \ref{fig:figA_cor_free_all}), making these reported abundances questionable. By contrast, our retrievals did not constrain H$_2$O and SiO abundances well, and statistical model comparisons confirmed their non-detection.

To investigate the of the discrepancies between the two works further, we compared the transmission spectra used in retrievals. The spectra in L25 exhibit significantly larger uncertainties than ours, despite showing a broadly similar structure, and are of much lower resolution. This may stem from their use of only a linear trend for NRS1 and no detrending for NRS2, in contrast to our more robust GP regression detrending applied to individual spectroscopic lightcurves (see Figure \ref{fig:figdetren}). Inflated uncertainties can indeed bias retrievals, as they may mimic molecular features. To test this, we ran retrievals on our polynomial-detrended spectra and found abundances consistent with our main results, albeit with larger uncertainties. Thus, inflated uncertainties alone do not explain the apparent discrepancies. We also noticed a systematic trend in the relative difference between the spectra reduced with the \texttt{FIREFLy} and \texttt{transitspectroscopy} pipelines in L25, which is large enough to drive small discrepancies, though we cannot provide a reasonable explanation for this.

Finally, comparing the retrieved PT profiles, we find the temperatures reported by L25 appear non-physically high. Given the strong irradiation of the day-side and the large day-night contrast in UHJs, upper atmospheric temperatures at terminators are expected to be much lower than the equilibrium temperature. PT profiles exceeding 3000K, as reported by L25, may therefore be superfluous. By contrast, our retrievals yielded consistent PT profiles across different datasets and parameterizations, which are also physically more plausible. Only the JWST-only isothermal PT profile from L25 is in close agreement with ours. Elevated temperature levels can significantly alter the retrieved abundances, as we also observed in our equilibrium chemistry retrievals when including H$_2$O.

Since H$_2$O and SiO are major oxygen carriers in UHJ atmospheres—apart from CO, which is well-constrained in our retrievals—we used the 3 $\sigma$ upper limits of both H$_2$O and SiO when estimating the C/O ratios. This approach avoids over estimating the ratio due to poorly constrained H$_2$O or SiO abundances. Using the retrieved abundances from the \texttt{Eureka!} spectrum with the Guillot PT profile, we estimate a C/O ratio of 0.954$\pm$0.033. This highly precise super-solar value highlights JWST's exceptional capabilities for constraining cross-molecular atmospheric properties. A super-solar C/O ratio suggests that WASP-178b likely formed beyond the ice line and migrated inward to its present orbit \citep{2017AJ....153...83B, 2021A&A...654A..71S}. However, material accretion during migration complicates efforts to distinguish between formation via core accretion and gravitational instability \citep{2014ApJ...794L..12M, 2024ApJ...969L..21B}. In addition, as suggested by \cite{2023MNRAS.520.4683F}, an elevated C/O ratio could also arise from oxygen depletion caused by sequestration into refractory condensates.

Interestingly, this super-solar C/O ratio is consistent with trends observed in other ultra-hot Jupiters \citep[e.g.,][]{2011Natur.469...64M, 2024AJ....168...14W, saha2025a, saha2025b}—in contrast to L25—suggesting a possible commonality in the physical and/or chemical processes shaping UHJ atmospheres. To explore a potential correlation between C/O ratio and equilibrium temperature, we compiled a sample of well-constrained C/O measurements spanning a broad range of equilibrium temperatures, as shown in Figure \ref{fig:figco}. Literature values were drawn from \citep{2024arXiv240303325B, 2023A&A...673A..98B, 2024AJ....168...14W, 2024ApJ...963L...5X, 2024ApJ...962L..30E, 2024AJ....167...43F, 2023MNRAS.525.2985R, 2023MNRAS.522.5062B, 2011Natur.469...64M, 2024A&A...685A..64K, 2025arXiv250601800W, 2025MNRAS.539.1381M, saha2025a, saha2025b}, with some equilibrium temperatures adopted from \citet{2023ApJS..268....2S, 2024ApJS..274...13S}. We modeled the observed C/O distribution as a quadratic function of equilibrium temperature: $ \mathrm{C/O} = \mathrm{a} \, \mathrm{T}_{\mathrm{eq}}^2 + \mathrm{b} \, \mathrm{T}_{\mathrm{eq}} + \mathrm{c} $, yielding best-fit coefficients of $\mathrm{a} = 0.72 \pm 0.25$, $\mathrm{b} = (-7.24 \pm 3.56) \times 10^{-4}$, and $\mathrm{c} = (3.25 \pm 1.11) \times 10^{-7}$. While the scatter remains significant and the current sample is limited, the fit indicates a possible trend. With JWST's ability to probe exoplanet atmospheres across a broad temperature range and deliver precise abundance measurements, extending such analyses to a broader population of UHJs and other exoplanet types will be critical for confirming this relation and for better disentangling the dominant physical and chemical processes at play in these extreme atmospheres. 

From the elevated abundances of CO in particular, we infer a high atmospheric metallicity of 11.44$_{-6.94}^{+12.54}$ $\times$solar. Such a metallicity is generally unexpected for Jupiter-sized gas-giants, where photoevaporation is typically thought to play only a minor role. However, this enrichment may reflect substantial atmospheric evolution, possibly driven by significant photoevaporation of hydrogen and helium in the upper atmosphere \citep[e.g.,][]{2018ApJ...865...75N, 2020arXiv200508676M}. Given that WASP-178b orbits a hot A-type star, intense high-energy irradiation could have enhanced this mass loss. Alternatively, the planet may have migrated inward during its early evolution \citep[e.g.,][]{2019A&A...627A.127C}, accreting large amounts of metal-rich dust from the inner protoplanetary disk \citep[e.g.,][]{1998AREPS..26...53B, 2021MNRAS.503.5254L, 2023A&A...675A.176U}, or possibly engulfing rocky planetesimals—or even smaller planets—during or after migration. Further JWST observations across a broader wavelength range—particularly day-side studies through emission spectroscopy or phase-curve analysis—will be critical for constraining the atmospheric properties and assessing the spatial distribution of this chemically enriched atmosphere.\linebreak\linebreak

SS acknowledges Fondo Comité Mixto-ESO Chile ORP 025/2022 to support this research. The computations presented in this work were performed using the Geryon-3 supercomputing cluster, which was assembled and maintained using funds provided by the ANID-BASAL Center FB210003, Center for Astrophysics and Associated Technologies, CATA. This work is based [in part] on observations made with the NASA/ESA/CSA James Webb Space Telescope. The data were obtained from the Mikulski Archive for Space Telescopes at the Space Telescope Science Institute, which is operated by the Association of Universities for Research in Astronomy, Inc., under NASA contract NAS 5-03127 for JWST. These observations are associated with program $\#$2055.  JSJ gratefully acknowledges support by FONDECYT grant 1240738 and from the ANID BASAL project FB210003.

\bibliography{ms}

\newpage

\section*{Appendix}

\renewcommand{\thefigure}{A\arabic{figure}}
\renewcommand{\thetable}{A\arabic{table}}
\renewcommand{\theequation}{A\arabic{equation}}
\renewcommand{\thepage}{A\arabic{page}}
\setcounter{figure}{0}
\setcounter{table}{0}
\setcounter{equation}{0}
\setcounter{page}{1}

\begin{figure*}[!h]
\centering
\begin{tabular}{@{}c@{}c@{}}
\includegraphics[width=0.5\columnwidth]{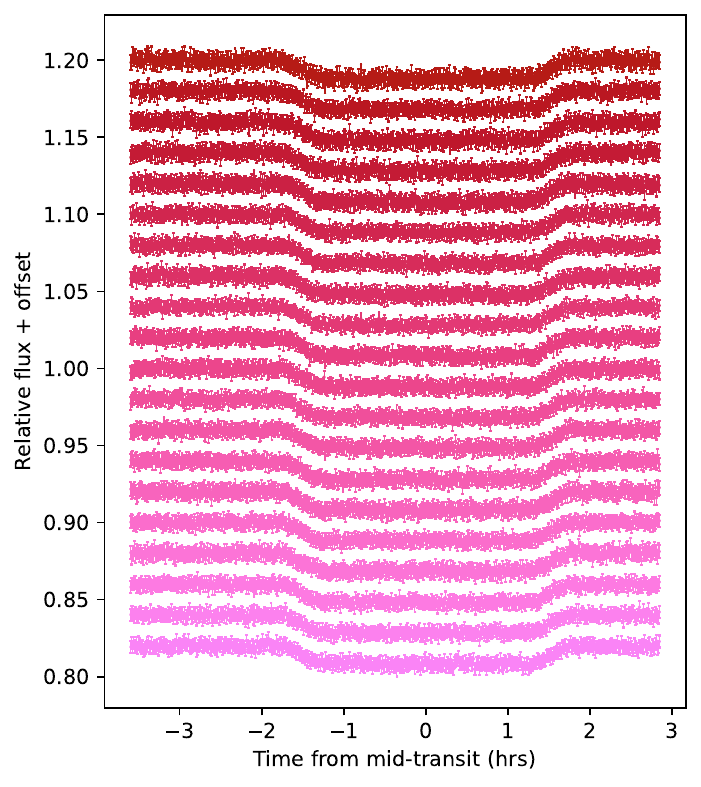} & \includegraphics[width=0.5\columnwidth]{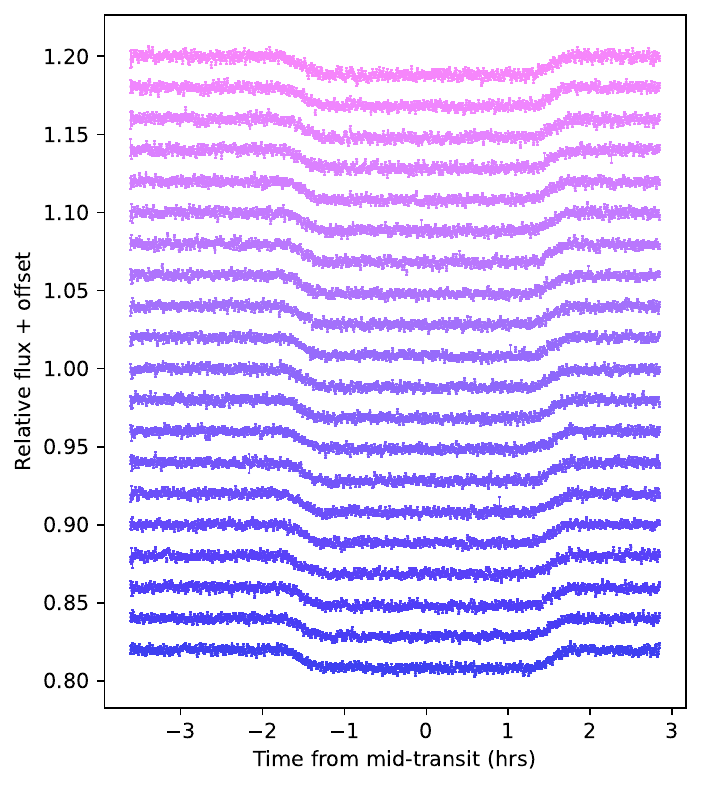}
\end{tabular}
\caption{The first 20 binned lightcurves from the \texttt{Eureka!} (left) and \texttt{exoTEDRF} (right) reductions. The \texttt{exoTEDRF} lightcurves exhibit visibly smaller error bars, as discussed in Section \ref{sec:sec2.3} and illustrated in Figure \ref{fig:figAerr}. \label{fig:figA1}}
\end{figure*}

\begin{figure*}[!h]
\centering
\begin{tabular}{@{}c@{}c@{}}
\includegraphics[width=0.5\columnwidth]{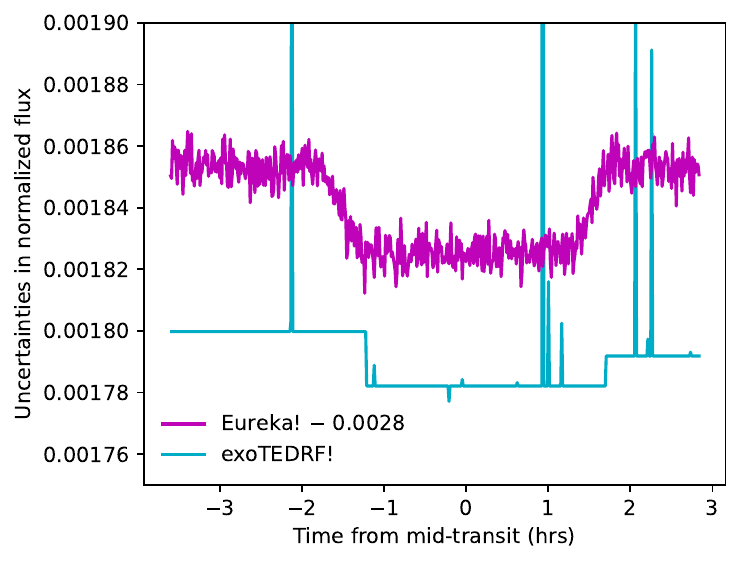} & \includegraphics[width=0.5\columnwidth]{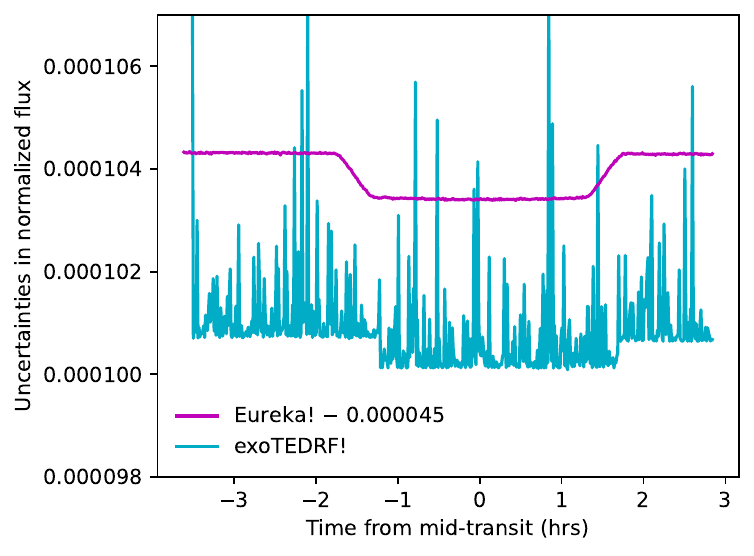}
\end{tabular}
\caption{The left panel compares the uncertainties in the first binned lightcurve from the \texttt{Eureka!} and \texttt{exoTEDRF} reductions, while the right panel shows the same for the NRS1 white lightcurves. The \texttt{exoTEDRF} uncertainties appear underestimated, exhibiting constant values across pixels but large inter-pixel fluctuations. However, in the white lightcurves, the fluctuations dominate the \texttt{exoTEDRF} uncertainties, thereby reducing the overall difference between the two pipelines. \label{fig:figAerr}}
\end{figure*}

\begin{table*}[!h]
    \centering
    \caption{Estimated parameters with their prior values from the white lightcurve fit (\texttt{Eureka!}). \label{tab:tr_fit}}
    \begin{tabular}{l c c}
        \hline
        \hline
        Parameter & Prior & Value \\
        \hline
        $P$ (days) & fixed & $3.3448369$ \\
        $t_0$ (MJD) & $\mathcal{U}(60010.41,60010.61)$ & $60010.50920 \pm 6.7 \times 10^{-5}$ \\
        $a/R_*$ & $\mathcal{U}(4.1,10.1)$ & $7.06 \pm 0.05$ \\
        $i$ (deg) & $\mathcal{U}(80,90)$ & $85.63 \pm 0.12$ \\
        $R_p/R_{*, \mathrm{NRS1}}$ & $\mathcal{U}(0.08,0.15)$ & $0.1074 \pm 0.0006$ \\
        $R_p/R_{*, \mathrm{NRS2}}$ & $\mathcal{U}(0.08,0.15)$ & $0.1075 \pm 0.0007$ \\
        $F_p/F_*$ & $\mathcal{U}(0,0.01)$ & $0.001725 \pm 1.5 \times 10^{-5}$ \\
        $e$ & fixed & $0$ \\
        $\omega$ (deg) & fixed & $90.0$ \\
        $u_{a, \mathrm{NRS1}}$ & $\mathcal{U}(0,1)$ & $0.089 \pm 0.046$ \\
        $u_{b, \mathrm{NRS1}}$ & $\mathcal{U}(0,1)$ & $0.045 \pm 0.070$ \\
        $u_{a, \mathrm{NRS2}}$ & $\mathcal{U}(0,1)$ & $0.052 \pm 0.049$ \\
        $u_{b, \mathrm{NRS2}}$ & $\mathcal{U}(0,1)$ & $0.066 \pm 0.064$ \\
        $f_{\mathrm{NRS1}}$ & $\mathcal{U}(0.998,1.002)$ & $0.9999 \pm 0.0009$ \\
        $f_{\mathrm{NRS2}}$ & $\mathcal{U}(0.998,1.002)$ & $1.0000 \pm 0.0003$ \\
        $\alpha_{\mathrm{NRS1}}$ & $\mathcal{LU}(-6,-1)$ & $-3.00 \pm 0.73$ \\
        $\tau_{\mathrm{NRS1}}$ & $\mathcal{LU}(-5,4)$ & $-0.16 \pm 0.65$ \\
        $\alpha_{\mathrm{NRS2}}$ & $\mathcal{LU}(-6,-1)$ & $-4.23 \pm 1.74$ \\
        $\tau_{\mathrm{NRS2}}$ & $\mathcal{LU}(-5,4)$ & $0.18 \pm 2.96$ \\
        \hline
        \hline
    \end{tabular}
\end{table*}

\begin{figure*}[!h]
\centering
\begin{tabular}{@{}c@{}c@{}}
\includegraphics[width=0.5\columnwidth]{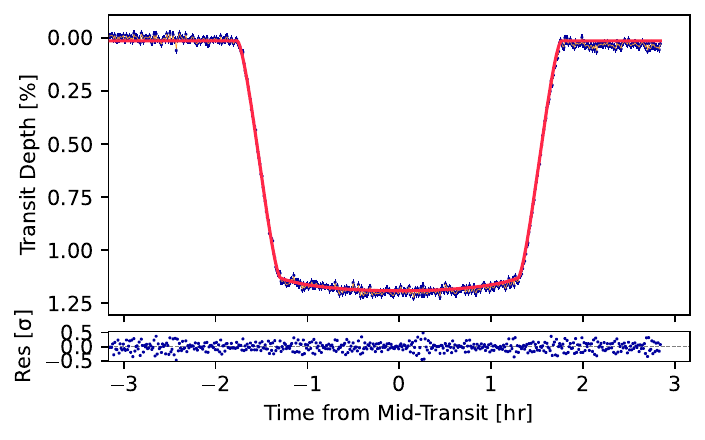} & \includegraphics[width=0.51\columnwidth]{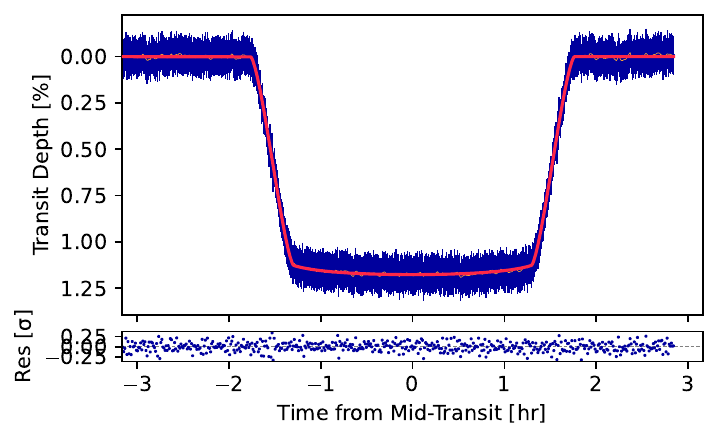}
\end{tabular}
\caption{NRS1 and NRS2 white lightcurves from the \texttt{Eureka!} reduction (blue), shown with the best fit transit model (red) and the combined transit plus GP model (orange thin line). Residuals are shown in units of observational uncertainties. \label{fig:fig_lc_w}}
\end{figure*}

\begin{figure*}[!h]
\includegraphics[width=1.\columnwidth]{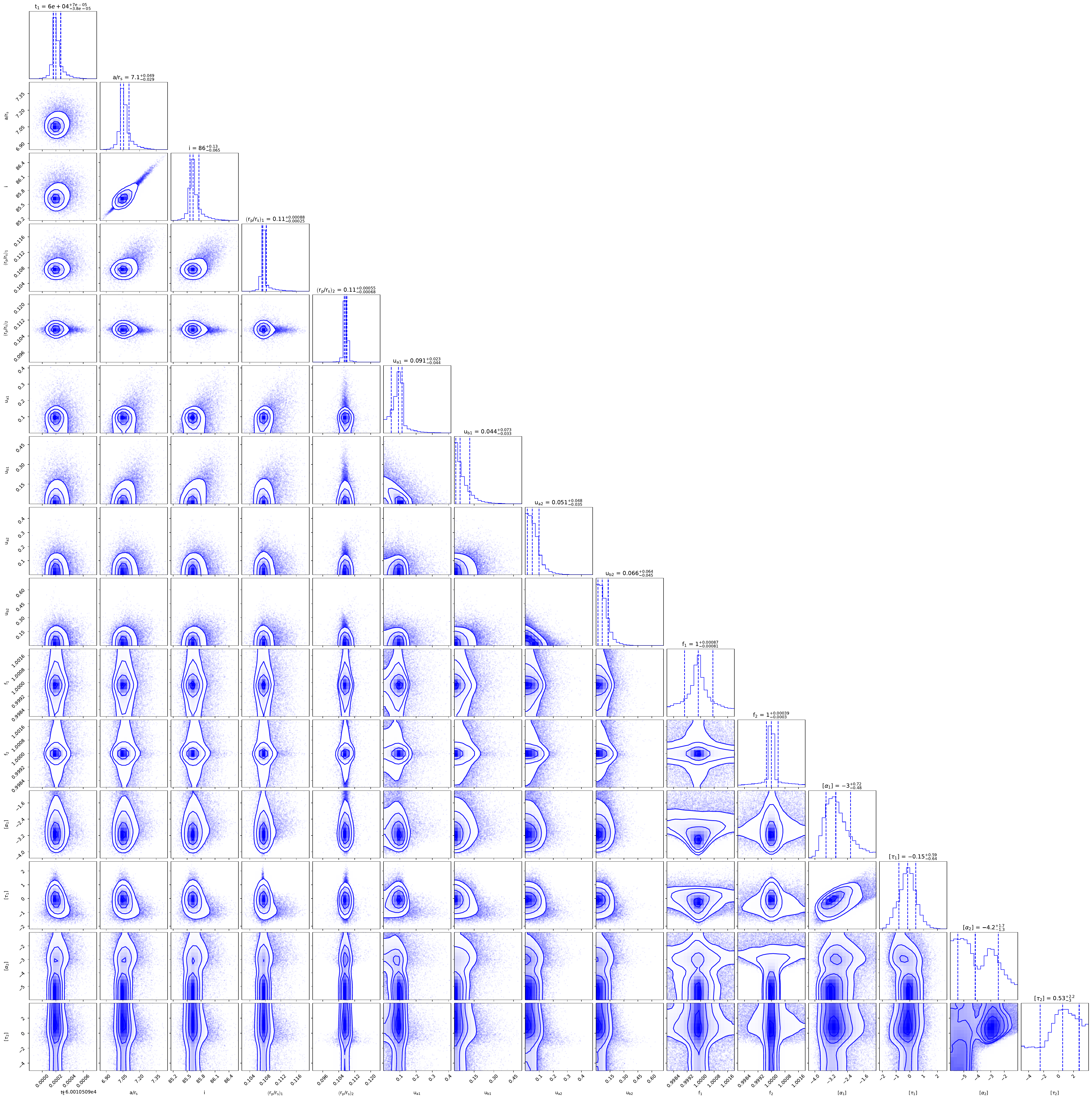}
\caption{Posterior distributions of the fitted parameters from the \texttt{Eureka!} white lightcurve fit. \label{fig:figA1b}}
\end{figure*}

\begin{figure}[!h]
\centering
\includegraphics[width=0.5\columnwidth]{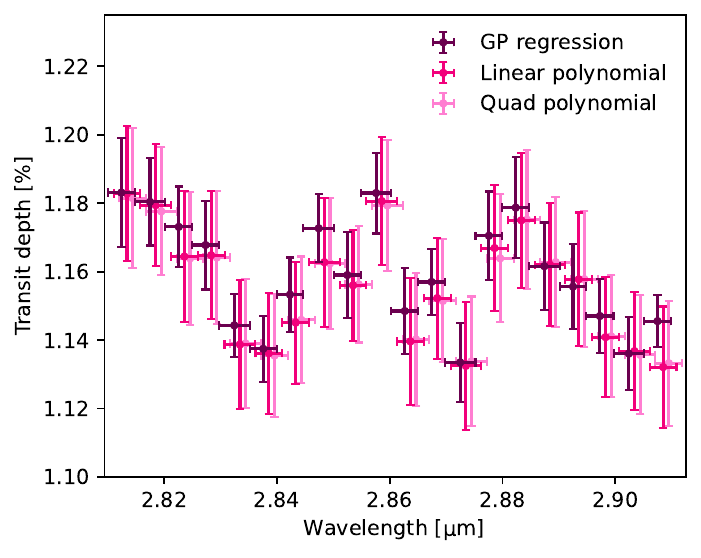}
\caption{Comparison of transmission spectra derived from the first 20 spectroscopic lightcurves from the \texttt{Eureka!} reduction, using different detrending algorithms in the lightcurve fitting. \label{fig:figdetren}}
\end{figure}

\begin{table*}[!h]
    \centering
    \caption{Prior distributions for the free parameters used in the free chemistry retrievals. \label{tab:retrieval_priors}}
    \begin{tabular}{l c}
        \hline
        \hline
        Parameter & Prior \\
        \hline
        $R_{\mathrm{p}}$ & $\mathcal{U}(1.6\,R_{\mathrm{Jup}}, 2.1\,R_{\mathrm{Jup}})$ \\
        $T_{\mathrm{equ}}$ & $\mathcal{U}(0, 5000)$ \\
        $T_{\mathrm{int}}$ & $\mathcal{U}(0, 500)$ \\
        $\gamma$ & $\mathcal{U}(0.01, 10)$ \\
        $[\kappa_{\mathrm{IR}}]$ & $\mathcal{U}(-4, 0)$ \\
        $T_{\mathrm{bottom}}$ & $\mathcal{U}(0, 5000)$ \\
        $s_{\mathrm{PT (1-10)}}$ & $\mathcal{U}(-0.5, 0.5)$ \\
        $f_{\mathrm{sed}}$ & $\mathcal{U}(0.1, 10)$ \\
        $\sigma_{\mathrm{lnorm}}$ & $\mathcal{U}(1.5, 3.0)$ \\
        $[K_{zz}]$ & $\mathcal{U}(2, 15)$ \\
        $f_c$ & $\mathcal{U}(0, 1)$ \\
        $f_{\mathrm{s, NRS2}}$ & $\mathcal{U}(0.95, 1.05)$ \\
        {[H$_2$O]} & $\mathcal{U}(-12, -0.2)$ \\
        {[CO]} & $\mathcal{U}(-12, -0.2)$ \\
        {[CO$_2$]} & $\mathcal{U}(-12, -0.2)$ \\
        {[SiO]} & $\mathcal{U}(-12, -0.2)$ \\
        {[C$_2$H$_2$]} & $\mathcal{U}(-12, -1.5)$ \\
        \hline
        \hline
    \end{tabular}
\end{table*}

\begin{figure*}[!h]
\includegraphics[width=1\columnwidth]{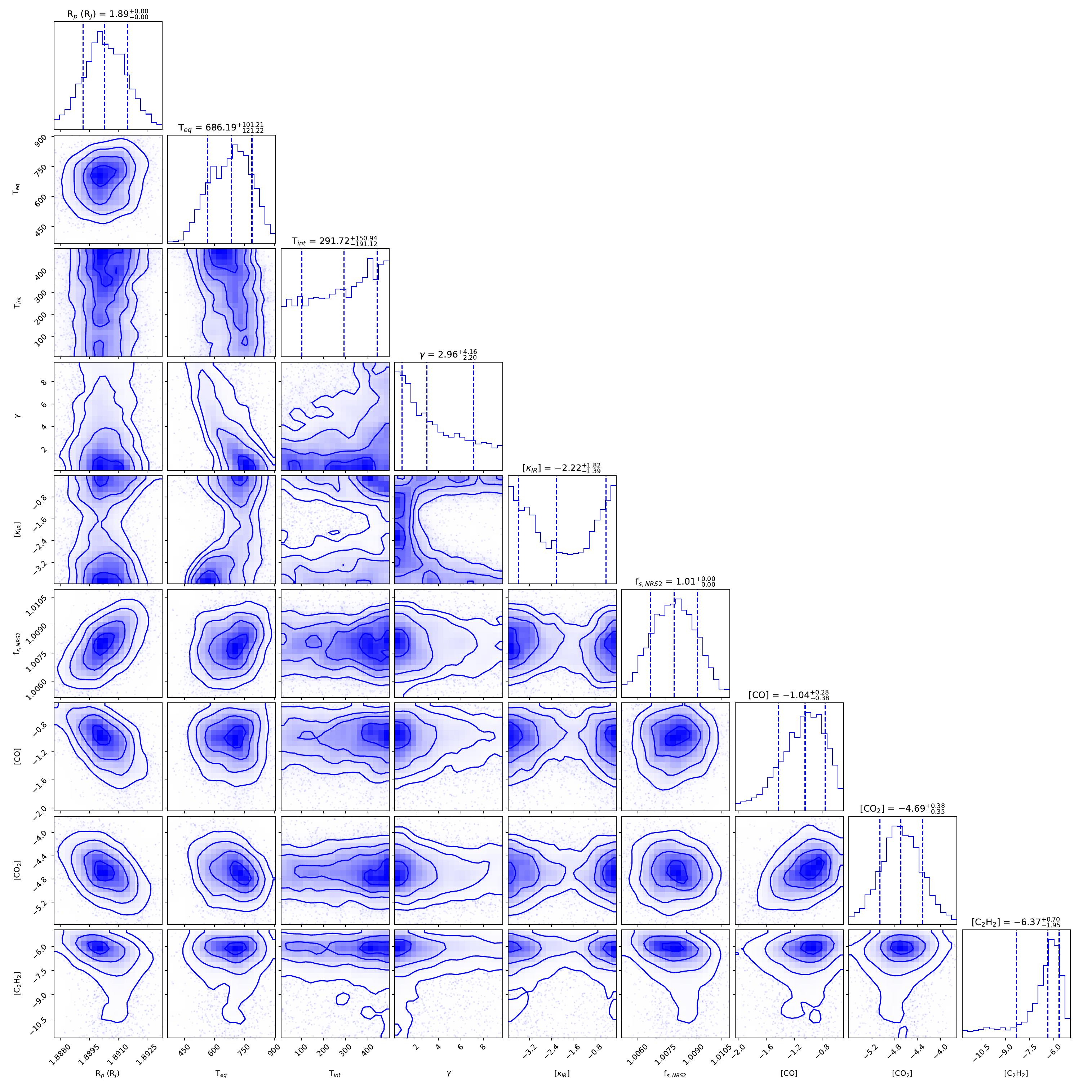}
\caption{Corner plot of the posterior distributions from the free chemistry retrieval of the \texttt{Eureka!} spectrum, including only the statistically detected molecular species. \label{fig:figA_cor_free}}
\end{figure*}

\begin{figure*}[!h]
\includegraphics[width=1\columnwidth]{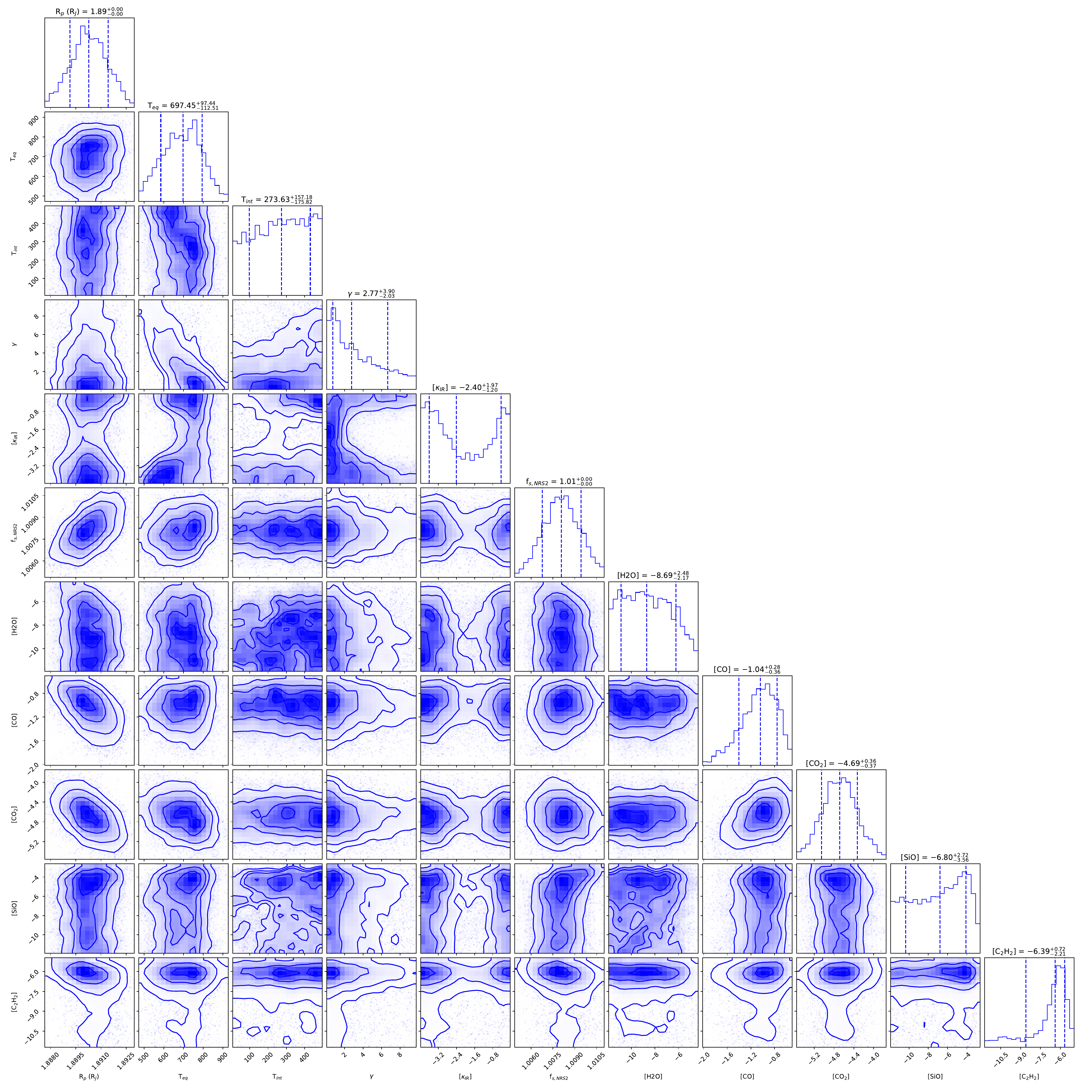}
\caption{Same as Figure \ref{fig:figA_cor_free}, but additionally including H$_2$O and SiO. \label{fig:figA_cor_free_all}}
\end{figure*}

\begin{figure*}[!h]
\includegraphics[width=1\columnwidth]{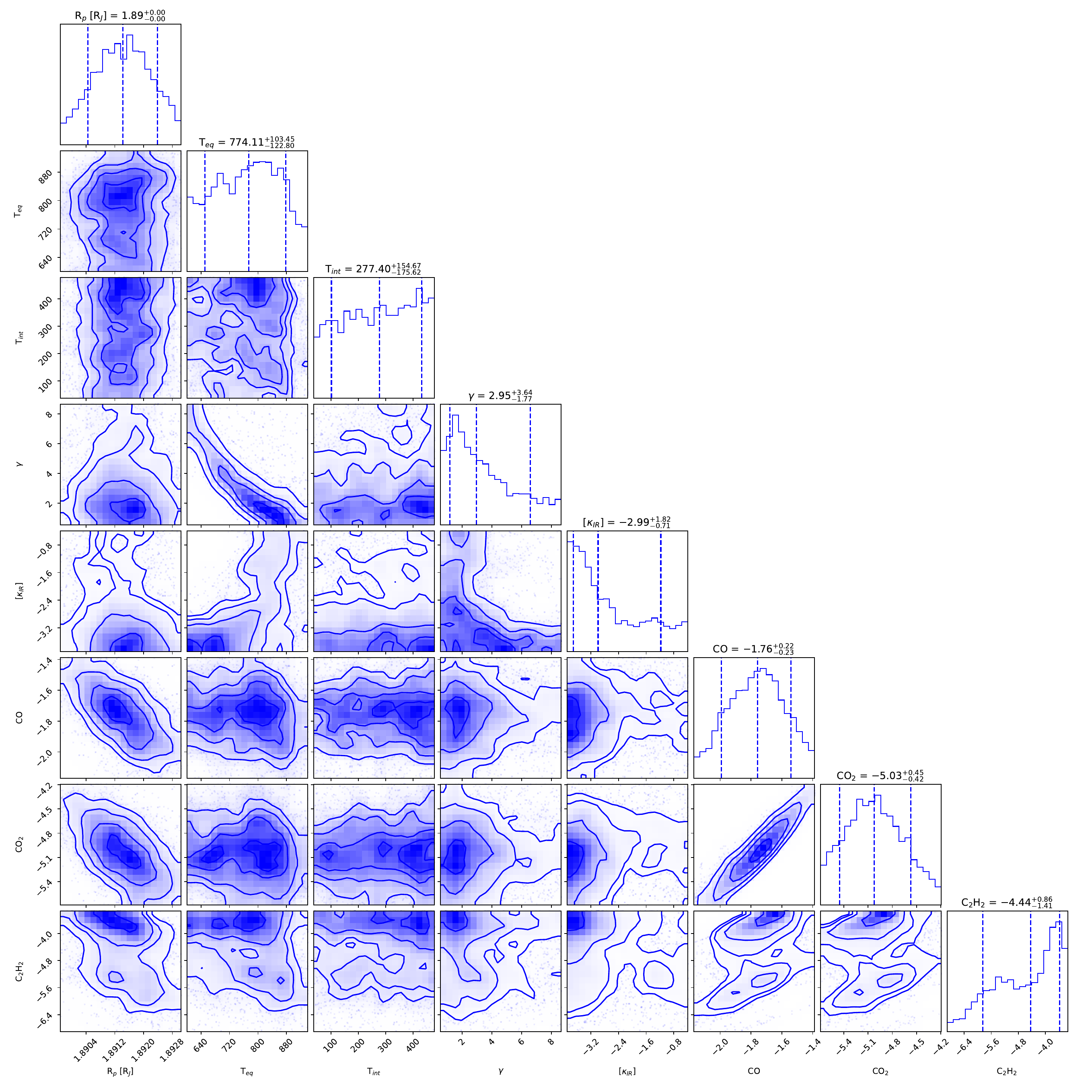}
\caption{Same as Figure \ref{fig:figA_cor_free}, but for the equilibrium chemistry retrieval. \label{fig:figA_cor_eq}}
\end{figure*}

\end{document}